\documentstyle[emulapj]{article}

\def\eps@scaling{1.0}
\def\epsscale#1{\gdef\eps@scaling{#1}}
\def\plotone#1{\centering \leavevmode
\epsfxsize=\eps@scaling\columnwidth \epsfbox{#1}}
\def\plottwo#1#2{\centering \leavevmode
\epsfxsize=\eps@scaling\columnwidth \epsfbox{#1} \hfil
\epsfxsize=\eps@scaling\columnwidth \epsfbox{#2}}

\begin{document}

\title{Origin of Blazar Activity}
\righthead{ROMANOVA}

\accepted{}

\medskip

\author{M.M. Romanova}
\affil {Space Research Institute of the Russian Academy of
 Sciences, Moscow, Russia; and
Department of Astronomy,
Cornell University, Ithaca, NY 14853--6801;
romanova@astrosun.tn.cornell.edu}

\slugcomment{For Proceedings of IAU Symposium No. 194,
Activity in Galaxies and Related Phenomena, August 17-21,
1998, Byurakan, Armenia}

\begin{abstract}

        Models of Blazars based on the propagation
of finite discontinuities or {\it fronts}
in the  Poynting flux jet
 from the innermost regions of an accretion  disk
around a black hole are discussed.
      Such fronts may be responsible for
short time--scale (from less than hours to days) flares
 in different wavebands from  high frequency radioband
to TeV, with  delay
in low radio frequencies as a result of
synchrotron self--absorption.
The cases of  magnetic fields of one and opposite
polarities across the front are investigated. We find
that annihilation of magnetic field in the front
leads to higher energy spectrum of leptons and
possibility of strong TeV flares.
Electron--positron pairs form in most cases
as a result of interaction between numerous
synchrotron photons and SSC photons,
and constitute the majority species, compared with
the ions at subparsec scales.
Frequent weak outbursts  may be responsible for
flickering core  radiation in all wavebands,
while the stronger outbursts may be observed as
short time--scale flares.

\end{abstract}


\section{Introduction}

   Blazars are characterized
by fast variability in different
wavebands from radio to gamma and in few cases by TeV
radiation (Urry \& Padovani 1995; Punch 1992).
   Many of the objects reveal ``superluminal'' jets,
which indicate that matter of the jet moves
nearly toward us  with
relativistic speed (Blandford \& Rees 1978).
   The variability of the objects may be connected
with outbursts of matter and energy from the
nucleus and propagation of
shocks along the collimated relativistic jet
(Blandford \& K\"onigl 1979; Marscher 1980).
   This model was further developed for
investigation of
different  radio properties of
radiogalaxies and quasars
(Aller, Aller \& Hughes 1985;
 Hughes, Aller \& Aller 1985).
   The back extrapolated time of VLBI  outbursts
approximately coincides with strong  optical,
X--ray and gamma--ray flares
(Kinman 1977;  Belokon  1988;
Krichbaum et al. 1995; Otterbein et al. 1998)
which is in favor of this model.
     The idea of matter outbursts from centers of
galaxies was first proposed  by Ambartsumian
(1958), and this has been confirmed  by
numerous observations.

   The outbursted matter may be a normal
electron--ion plasma, or it may consist
mainly of electron--positron pairs.
   Prior to the Compton Observatory measurements,
prediction of strong, collimated
gamma--ray emission and electron/positron cascades in AGN
relativistic jets  was made by the
 model of Lovelace, MacAuslan \&
Burns (1979); Burns \& Lovelace (1982).
    More recently, a number of theoretical
models have been developed to explain
the observed gamma--ray emission of AGNs.
    In most of the models the gamma--ray
radiation is ascribed to inverse Compton (IC)
scattering of relativistic electrons
and possibly positrons (Lorentz factors
$\gamma \sim 10^2-10^5$) of a jet having
relativistic bulk motion
 (Lorentz factor $\Gamma \sim 10$)
with soft photons (energies $\sim 1-10^2$ eV).
     The soft photons can
arise from the synchrotron emission
of the relativistic
electrons in the jet as in the
synchrotron--self--Compton
(SSC) models (Maraschi, Ghisellini
\& Celotti 1992; Marscher \& Bloom 1992),
or from the direct
or scattered thermal radiation from an
accretion disk (Dermer, Schlickeiser \&
Mastichiadis 1992; Blandford 1993; Sikora,
Begelman \& Rees 1994),
or from a single cloud (Ghisellini \& Madau 1996).
   In a very  different class of models, ultra
high--energy protons (Lorentz factors $>10^6$) are
postulated to cause a cascade,
the product particles
of which produce the observed radiation
(Mannheim \& Biermann 1992;
Protheroe \& Biermann 1997).

    The idea of Poynting flux outbursts of energy
to the jet is based on the fact that
the central regions of the disk and a black
 hole may accumulate strong poloidal magnetic field
of the order $B\sim (10^3 - 10^4)~{\rm G}$.
    Rotation of this
configuration leads to generation of the Poynting
flux, which is a  ``permanent machine''
for matter acceleration (Blandford \& Znajek 1977;
Lovelace, Wang \& Sulkanen 1987; Livio,
 Ogilvio \& Pringle 1998).
     Recently, this idea was further developed and
applied to gamma--ray Blazars by Romanova \& Lovelace
(1997) (hereafter RL97), Colgate \& Li (1998) and
by  Levinson (1998).
     RL97 proposed that the main driving
force for the observed superluminal jet
components is a finite amplitude discontinuity
in a Poynting flux jet.
   A rapid change in the
Poynting jet outflow from a
disk can result from implosive accretion in a disk
with an ordered magnetic field
(Lovelace, Romanova \& Newman 1994, hereafter LRN94).
    Propagation of
newly expelled electromagnetic field and matter
from the disk with higher velocity than
the old jet
can  lead to
the formation of a pair of shock
waves as in the  non--relativistic
hydrodynamic flows in optical
jet in protostellar systems (Raga et al. 1990).
   Particle acceleration in the front may
result from the shocks and/or from
annihilation and reconnection of
oppositely directed magnetic
fields in the front (Romanova \& Lovelace 1992, hereafter RL92;
 Lovelace, Newman \& Romanova 1997, hereafter LNR97).
   Here, we consider
the different aspects of the flares of
 Blazars interpreted in terms
of discontinuities
in a Poynting flux  jet.
    We analyze different theoretical and
observational aspects connected
with such outbursts.

\begin{figure*}[t]
\epsscale{1.0}
\plotone{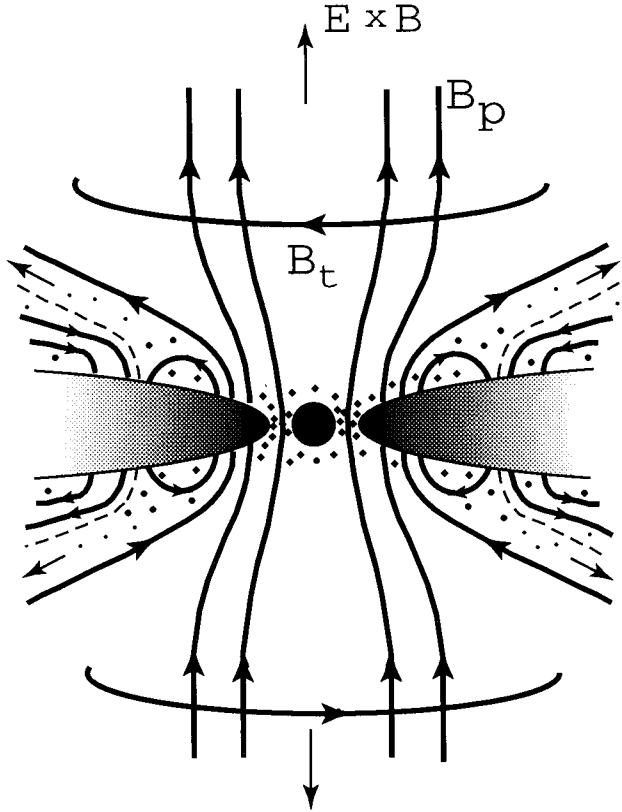}
\caption{Sketch of magnetic field threading the
 inner region of an accretion  disk around a black hole.
    A strong flux of electromagnetic energy (Poynting flux)
forms as a result of twisting of the poloidal component of
the magnetic field, and the
 fact that ${\bf E}=-{\bf v}\times{\bf B}/c$
in highly conducting plasma.}
\end{figure*}

\section{Magnetic Field in Jets}

       Accreting matter in a disk around
a   black hole carries with it an ordered and
a chaotic magnetic field.
        When the matter reaches the black hole,
small magnetic loops
 reconnect and the field annihilates.
     The more ordered
component of the field, which can
have open field lines, can
be dragged into the black hole while
remaining  connected to more distant
plasma in the corona of the disk (
Figure 1).
      Thus, the black hole may have a significant
magnetic field passing through it
supported by  external currents in the disk
and corona (Blandford \& Znajek 1977;
Macdonald \& Thorne 1982).
        The magnetic field of the
inner regions of the disk
is likely to be comparable to the field
 in the black hole
(Macdonald \& Thorne 1982; Livio et al. 1998).
      The area of the inner region of the disk
is much larger than the area of the black hole,
so that  the magnetic flux and the Poynting
energy outflow rate of the inner regions of the
 disk is larger than that of the black hole
(Livio et al. 1998;
Lovelace et al. 1987).
       The magnetic field near the
 inner edge of the disk is deduced to be of the order  of
$B\sim (10^3-10^4)~{\rm G}$ (Lovelace 1976).
     Semi--empirical models of gamma--ray flares
based on inverse Compton and/or SSC mechanisms,
predict approximate values of the magnetic field
in  different Blazars in the regions of origin of
the radiation
(e.g., Sambruna, Maraschi \& Urry 1996;
Sambruna et al. 1997).
     Back extrapolated to the inner disk
using $B \sim 1/r_j$ ($r_j(z)$ is the jet
radius), gives a  magnetic field $B=(10^2-10^4)~{\rm G}$.

    The rotating disk and black hole threaded by
an ordered magnetic field generate Poynting flux
outflows, or jets in which the energy density
of the electromagnetic field is much larger
than the matter energy density (see Lovelace,
these Proceedings).

\section{ Poynting Flux  Jets}

   Poynting flux winds were
first discussed by Goldreich and Julian (1968) for pulsars,
 and
later Poynting flux jets were  proposed to explain
extragalactic jets by
Lovelace (1976) and Blandford (1976).
  Solutions for Poynting flux outflows from
a disk around a massive black hole were
investigated by Lovelace et al.  (1987).

   A Poynting flux jet is self--collimated,
with energy, momentum, and angular momentum transported mainly
by the electromagnetic field (Lovelace et al. 1987).
   The collimation is due to the toroidal component of
magnetic field.
   A steady Poynting flux jet is characterized
in the lab frame by its asymptotic  ( $ z >> r_o $ )  magnetic
field $B_\phi = - B_0 [r_0/r_j(z)]$ and electric field
$E_r = - (v_j/c) B_0 [r_0/r_j(z)]$ at the jet's edge
with radius $r = r_j(z)$, where $r_0$
is the jet's radius at $z = 0$. Also,
$B_0$ is magnetic field   at $z = 0$, and $v_j \approx c$
is jet's axial velocity.
  The  jet radius at $z=0$
is $r_0 \sim (1-3) r_g = (2-6) GM/c^2$, where
$r_g$ is the Schwarzschild radius, and $M$ is the
black hole mass.
   The energy flux (luminosity) of the $+z$
jet is
$$
L_j = {v_j B_0^2 r_0^2/8}
\approx {3.0\times 10^{43}
~{\rm erg~s}^{-1}}~
(v_j / c) B_3^2 M_8^2,
\eqno(1)
$$
where $B_3\equiv {B_0/{10^3~{\rm G}}}$,
$M_8\equiv M/{10^8 M_\odot}$, and $r_0=3 r_g$.

Matter accreting in the disk will be
partially expelled to the jet by the Poynting flux
``machine.''
   An important quantity is the ratio of the
magnetic energy to the
rest mass energy at the base of the jet
$$
\mu \equiv (B_0^2/8\pi)/(\rho_0 c^2)~.
\eqno(2)
$$
   The Poynting regime corresponds to $\mu >> 1$.
Note that in the Poynting flux regime, the
magnetic field lines do not need to be inclined away
from the $z-$axis in order for there to be
energy outflow from the disk.
   This is in marked contrast with the hydromagnetic
regime where the outflows require the field to be inclined
away from the $z-$axis by $>30^o$
(Blandford \& Payne 1982; Lovelace, Berk \& Contopoulos 1991;
Romanova et al. 1997).

The matter flux carried by the jet  is
$$
\dot M_j = \pi r_0^2 \rho_0 v_{j} \approx {{B_0^2 r_0^2}
/{(8 \mu c)}}
$$
$$
\approx {5.2\times 10^{-4}}~
{{\rm  M_\odot}{\rm yr}^{-1}}~
{B_3^2 M_8^2/\mu}.
\eqno(3)
$$
At say $\mu=10$, the matter flux is $\dot M_j=5.2\times 10^{-5}
~{{\rm  M_\odot}/ {\rm yr}}$,
which is quite small compared with the disk accretion
rate needed to fuel an AGN, $\dot{M}_{accr} \sim M_\odot/$yr.

\subsection{Non--stationary Accretion and  Formation of
Fronts}

  The electromagnetic energy outflow from
the inner part of the accretion disk
and from the black hole may be
relatively steady if the disk
accretion flow is steady.
  There may of course be small inhomogeneities
connected with changes in the
density and magnetic field
of the accreting matter.
   However, if the disk is unstable, then the
accretion rate increases
and magnetic field also increases
as a result of matter compression.

    Inhomogeneity of the  magnetic field
threading the
disk can lead to a
``global magnetic instability"
of the accretion disk, which is
connected with
angular momentum outflow to Poynting
flux jets or hydromagnetic outflows
(LRN94;
Lubow, Papaloizou \& Pringle 1994;
 LNR97).
   If the  magnetic field is enhanced at
some radius $r$
in the disk, then angular momentum will be
the more efficiently lost from this region of the
disk's surfaces to the jets.
   As a result, this disk matter will accrete
faster and will accumulate matter in front of it
as it moves radially inward.
  This will in turn amplify the
magnetic field and further
increase the loss of angular momentum to jets.
    Simulations
of this process have shown that a strong
wave--pulse of dense
matter with strong magnetic field  forms
 in the disk and propagates
inward (LRN94).
  In case of a relatively high turbulent
or ``$\alpha$'' viscosity,
a soliton--like wave forms (LRN94), while in case
of a zero viscosity disk, a
shock--like wave forms (LNR97).
   When this wave
reaches the inner part of the disk, it generates the strong
outburst of energy, angular momentum and matter to the jet.
   The pulse brings in a stronger magnetic field
to the inner regions of the jet, which increases the
Poynting flux.
  Thus, the new jet outflow
will have a larger
magnetic field, and can
form a {\it front} propagating outward
along the channel of the ``old",  weaker jet.
  Also, the new jet outflow may involve
a reversal of polarity of the magnetic field
(Romanova et al. 1998)
so that the toroidal fields
of the ``new" and ``old" jets
are opposite (LNR97).

\subsection{Particle Acceleration in
Propagating Front of Jet}

  The  plasma newly launched to
the Poynting flux jet
propagates along the
old channel of the jet forming a {\it front}.
  The front is
bounded by two shock waves -- the
first between the front and
upstream matter and the second between the front
and the downstream matter.
  Parameters of matter
in the front are those in between
the ``old'' and ``new" jets.
  The difference in the  energy fluxes between
the ``old'' and ``new'' jets
necessarily goes into the acceleration of  particles
inside the front (RL97).
   It is  not possible from first principles to
calculate the spectrum of accelerated particles because
the acceleration mechanism is not known.
  Relativistic MHD shock acceleration may
be important (Eilek \& Hughes 1990), but
this is likely to accelerate mainly the
ions if the plasma consists mainly of electrons
and ions.
  From the other side, if the toroidal
magnetic field reverses polarity
across the front, then
reconnection of magnetic field may be the
dominant particle acceleration mechanism.
    Also, both particle acceleration mechanisms
may be significantly different if the plasma consists
mainly of electrons and positrons.
   Figure 2 shows sketch of possible configurations
of the magnetic field in the front.

\begin{figure*}[t]
\epsscale{1.0}
\plotone{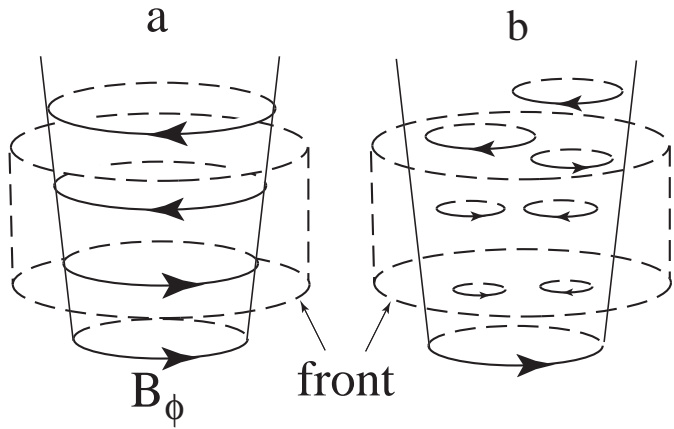}
\caption{
Sketches of possible  configurations
of magnetic field with opposite polarity,
(a) an ordered field and (b) a chaotic
 field.}
\end{figure*}

If accreting matter in the disk carries poloidal
magnetic field of both polarities
this can lead to reversals in the polarity
of the toroidal magnetic field of the Poynting
jet (see LNR97).
  This situation is sketched in
Figure 2a.
   The case of a
chaotic magnetic field (Figure 2b) may be always
present in the plasma.
   Compression of plasma will lead to
driven collisionless reconnection
(Alfv\'en 1968; RL92).

RL97 proposed an empirical lepton spectrum
motivated by observations where the energy of
leptons is distributed as $f_{l} \sim \gamma^{-2}$ between
 energies $\gamma_1$, and $\gamma_2$ (in units of
the rest mass energy), and
is steeper $\sim \gamma^{-3}$ between energies
$\gamma_2$ and $\gamma_3$.
   The lowest energy $\gamma_1$
was equal to that corresponding to
synchrotron self--absorption.
   The average energy $\bar \gamma$
was calculated from the basic equations for
the conservation of energy, mass and momentum.
   The energies $\gamma_2$, and $\gamma_3$ were derived
from other considerations (see RL97).
   The actual spectrum $f_l$
may be somewhat different and
will be determined by details of the acceleration
process(es).

  The basic equations for the evolution of
the front follow from the conservation of mass,
momentum, and energy,
$$
{{\partial n_i'}\over{\partial t'}} =
-{{\partial (n_i'v')}\over {\partial z'}} +
{\Big ({{\partial n_i'}\over{\partial t'}}\Big )}_{pairs},
\eqno(4)
$$
$$
{{\partial T_{0z}'}\over {\partial t'}}=
-{{\partial T_{zz}'}\over{\partial z'}} + grav + rad~,
\eqno(5)
$$
$$
{{\partial T_{oo}'}\over {\partial t'}}
= - {{\partial T_{oz}'}\over {\partial z'}}
-syn - SSC - Com,
\eqno(6)
$$
and an equation for the magnetic flux.
Here, $T_{\alpha \beta}^\prime $ is the stress energy
tensor of the matter and the electromagnetic
field, the primes denote quantities in the
frame of the front, and ``grav,'' ``syn,'' etc
denote different forces or energy fluxes
discussed by RL97.

The equations were solved for different intrinsic
parameters (RL97),
but the example used below corresponds to a
black hole mass $M = 3\times 10^8 M_{\odot}$,
 magnetic field at the base
of the jet $B_0 = 10^3~{\rm G}$,
initial ion/lepton ratio $f_{li}=1$,
magnetic/particle energy
at the base of the jet $\mu=15$,
 viewing angle $\theta=0.2~{\rm rad}$, the
luminosity of background photons
 $L_{ph}=10^{46} {\rm erg~s^{-1}}$, their energy
$\epsilon_{ph}=10~ {\rm eV}$, and radius of
their  distribution
$R_{ph}\approx 3\times 10^{17} {\rm cm}$.
The density and magnetic field ratios between
the ``old" and ``new" matter
are $n_1/n_2 \approx 0.4$ and $B_1/B_2 \approx 0.4$.
      The velocities correspond to a
bulk Lorentz factors $\Gamma_1=8$ and $\Gamma_2=18$
(where $\Gamma=[1-(v_j/c)^2]^{-1/2}$).
The cases of  a single and reverse  polarity
of the magnetic field across the front were
investigated and compared.
After expulsion of new matter,
the front accelerates up to
$\Gamma\approx 12$, so that
the Doppler boost factor
is $\delta =
{1/{\Gamma[1-(v_j/c) {\rm cos} \theta]}} \approx 4$.
Leptons are accelerated in the front from $\gamma=1$
 to $\gamma_1\approx 10^2$,
 $\gamma_2\approx 10^3-5\times 10^3$,
and $\gamma_3\approx (6\times 10^3-2\times 10^4)$.
In the  case of reversal polarity
leptons are accelerated up to higher energies:
$\gamma_2\approx 10^4-10^5$, and
$\gamma_3\approx 10^5-10^6$.
 These maximum values of $\gamma$ depend
on the duration of expulsion of ``new" matter
with opposite polarity.

\subsection{Energy Release Due to Magnetic Field
Reconnection}

  Reconnection of the magnetic field
may occur  along the jet, in particular, in the front,
where matter is compressed.
   The magnetic field at large distances
is $B_\phi = - B_0 (r_0/r_j)$.
  The magnetic energy--density at
some distance $z$ along the jet is  $B_\phi^2/{8 \pi}$
so that the total magnetic
energy in the front is $E_m=(B_\phi^2/{8\pi})(\pi r^3)$,
where we supposed that the region is a
cylinder with a length
equal to its radius.
  If the magnetic field reverses polarity across
the front,
then this magnetic energy may be
released entirely in the form
of accelerated particles during
an Alfv\'en time $t_A = r_j/v_A$,
where the  Alfv\'en velocity
$v_A = c/(1+ 4\pi \rho c^2/B_\phi^2)^{1/2} \sim c$.
Thus, the ``luminosity''
owing to the reconnection is
$$
  L_{\rm recon} \sim {E_m / t_A}=
{B_0^2 r_0^2 c}/8 \sim 3.0\times 10^{43}
{\rm erg~s}^{-1}~ B_3^2 M_8^2.
\eqno(7)
$$
The spectrum of leptons resulting from
collisionless driven reconnection is a power law
$\gamma^{-2}$ for electron-ion plasma, and
$\gamma^{-1.5}$ for electron-positron plasma
(RL92).
In our analysis of the time evolution
of the front in a Poynting flux jet,
we took into account the annihilation of magnetic
 field in the case where the field reverses
polarity across the front.
   We found
larger   particles energies in
the case where the field reverses polarity.

\subsection{Pair Creation}

  The particle content of the jets is
not known (Krolik, these Proceedings).
In our model,
the density of synchrotron photons inside the front
is  typically much larger
($10^4-10^5$ times)
 than the density of the background
photons (see also RL97).
   Interaction of the electrons
with these photons produces a
high density of high--energy SSC photons.
Analysis of different possible mechanisms of pair
creation leads to the conclusion that interaction
of SSC photons with synchrotron
photons is the most important
process.
  A pair forms
when $\epsilon_{syn} \epsilon_{ssc} > (m_e c^2)^2$,
where $\epsilon_{syn}$ and $\epsilon_{ssc}$
are energies of synchrotron and SSC photons.
  For rough estimates, we can write
$\epsilon_{syn}=(3/2) \gamma^2 \hbar
\omega_o'$, where $\omega_o' = e| B'|/(m_ec)$ is
the cyclotron frequency in the front frame,
$\epsilon_{ssc} = \gamma^2 \epsilon_{syn}$, and $|B'|$
is the magnetic field strength in the front
frame. An approximate condition for
pair production is
$$
 \gamma \geq \gamma_{pair}
\equiv \big(mc^2/\hbar\omega_o'\big)^{1/3}
\approx 3.5\times 10^3 (|B'_3|)^{-1/3}.
\eqno(8)
$$
    Electron--positron recombination is negligible
for the conditions considered.
We observed that the pairs form
at a variety of parameters
of the model.  In a typical case, the total number of
pairs in the front $N_l$
grows  proportionally to the total number of ions
$N_i$  accumulated by the front.
   Thus, their
ratio $f_{li}\equiv N_l/N_i$
is almost constant during
the front propagation and constitutes
 $f_{li}\approx 10$ for $\mu=15$, and
 $f_{li}\approx 100$ for $\mu=100$.
   The pair creation does not appreciably
alter the energy distribution of leptons because
the  internal energy per particle
is proportional to $\mu$ and
the number of leptons is
approximately proportional to $\mu$.
   In RL97, pairs did not form, because at similar
initial  parameters, they took initial value
 $f_{li}=5$ at the base of the jet.
   Thus, leptons
had smaller initial energy per particle
and the condition $\gamma > \gamma_{pair}$
was not  satisfied.
   Summarizing, we conclude that  pairs form in many
cases, but
only in case of very low matter energy compared to
magnetic energy, $\mu >>1$,
their number may be significant
compared to the number of ions.

In case of polarity reversals of the
$B_\phi$ field (Figure 2), the
magnetic field gradually
annihilates and becomes weaker in the
front frame.
 The pairs form in the
beginning of the front
propagation ($z < 10^{16} M_8~{\rm cm}$),
 when the magnetic field
is strong enough  to produce
significant population of
 synchrotron and SSC photons.
   Later, the rate of pair creation
decreases, and $f_{li}$ decreases.
   The fact, that in
 case of polarity reversals, the internal energy
of  the front is larger
(due to reconnection) while the number of leptons
is smaller (due to weaker pair creation)
leads to higher
 energies per lepton $\gamma$ and the
entire spectrum is harder.

\subsection{Radiation from the Front}

  Accelerated leptons interact with
background and synchrotron photons.
   Each of three processes, synchrotron, IC, and SSC,
give  radiation in a broad bands of energies.
   They cover the whole range of
energies from
$10^{-3}~{\rm eV}$
(radio with $\lambda\approx 1~{\rm mm}$) up
 to  $\sim 100$ GeV in case of single polarity and up to
 $(1-10)~{\rm TeV}$ in case of polarity reversals.
   Thus, the Poynting flux outburst generates
a simultaneous flare in all wavebands
 from high frequency radio
up to high energy radiation.
Here, we calculated the luminosity of flares in EGRET band
$(30~{\rm MeV}-30~{\rm GeV})$ and
in TeV band for energies $(0.1-10)~{\rm TeV}$
(see Figure 3).
   The cases of one polarity (1)
and two polarities (2) of the toroidal magnetic
field  are shown.
\begin{figure*}[t]
\epsscale{1.0}
\plottwo{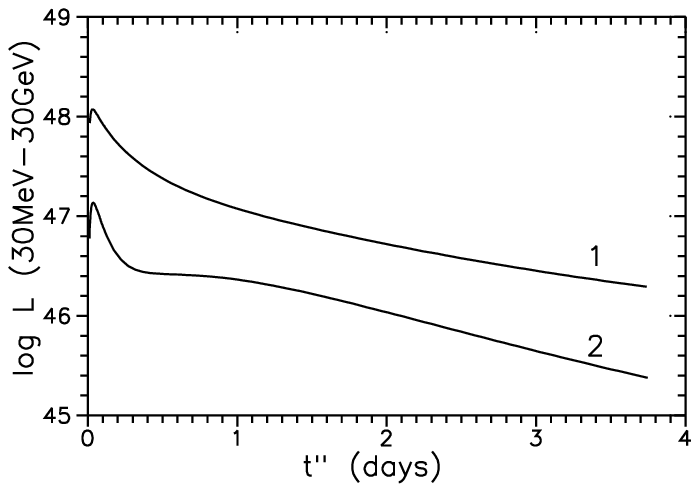}{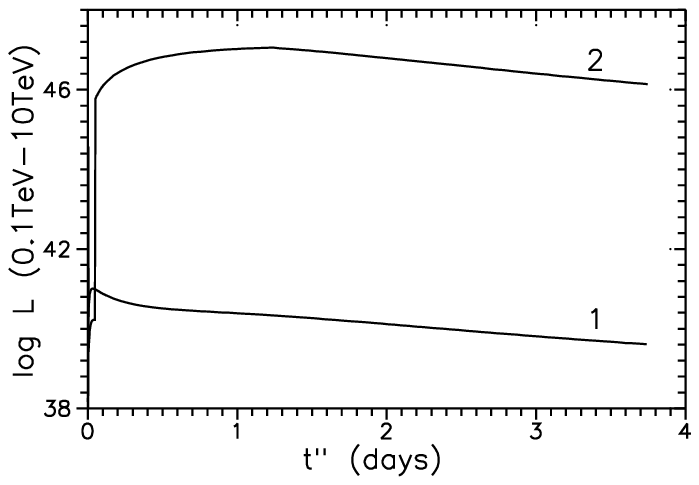}
\caption{
Luminosity of EGRET flares
 (left panel) and TeV flares
 (right panel)
in ${\rm erg/s}$
 for cases of (1) a magnetic field of
one polarity, and (2) a field which reverses
polarity across the front.}
\end{figure*}
   One can see that the gamma--ray luminosity
is higher in the case of a
single polarity, whereas the
TeV luminosity is much higher in the case of
reversals of polarity of
the field across the front.
   In case of one polarity,
radiation in both energy bands
is determined by SSC processes,
while in case of polarity reversal,
both gamma and TeV
radiation  are determined mainly by
the IC processes (because the magnetic field
becomes too weak to produce significant synchrotron
and correspondingly SSC radiation).
  The fairly strong TeV flare in case of polarity
reversals appears as a result of higher energy
of leptons in the front compared with the case
of a single polarity.
  The flare gradually decreases because the density of
background photons decreases.
  However, the strong luminosity
flare may appear later in the case of interaction with
a Broad Line Region cloud (Ghisellini \& Madau 1996).
   The flares in X--ray and optical bands are smaller,
than in the EGRET band.
The  detailed analysis of the spectrum
will be considered in the future.

    In the example shown (Figure 3) it was suggested
that the ``new'' matter was outbursted
from the disk during $\sim 100$ periods
 of rotation of the inner
radius of the disk, $\Delta t = 100 (2\pi r_0/v_{K0})$,
  which corresponds to
 $\Delta t'' \approx 4~{\rm days}$
 in the observer's frame
 ($dt''=  dt(\Gamma/\delta) =
dt[1-(v_j/c) {\rm cos}\theta])$.
     If the source of new matter ceases earlier
(see also Levinson 1998),
then the flare may be much shorter, of
the order of the dynamical time--scale at the inner
edge of the disk,
$\Delta t''_{dyn}= t_{dyn}(\Gamma/\delta) \approx
0.8 (\Gamma/\delta)M_8~{\rm hr} \approx 2.4M_8~{\rm hr}$.
  Thus, the flares shown at Figure 3, may be much
shorter.
 The discussed model presents definite answer
regarding the amplitude of the flares, but their duration
is somewhat uncertain and is determined by the
processes in the inner regions of the disk.
   We expect that the inner regions of the disk
may frequently be unstable owing to angular momentum
outflow  to the jet.
  This is an analogue of
the ``global magnetic instability" (LRN94; LNR97), but
on smaller scales. This will lead to frequent
outbursts of matter to the jet with characteristic time,
which may be as short as few dynamical time--scales.
   One can expect numerous
hour time--scale flares with less frequent,
 more powerful, day time--scale flares
determined by stronger
instabilities in the disk.

\section{Comparison with Observations}

  In Blazars
 we are looking down the jet and
this gives information not
visible in other
AGNs (Sambruna, these Proceedings).
   But {\it how deep can one observe the
inner jet}?
       The interaction of gamma--ray photons with
few KeV background radiation leads to electron--positron
pair production (e.g., Burns \& Lovelace 1982;
Blandford \& Levinson 1995).
     The corresponding $\gamma-$sphere
has a radius (Takahara 1997):
$r_\gamma > 1.0\times10^{-5}
{(\epsilon_{{\rm ob}}/{\rm GeV})}
L_{46} \delta_{10}^{-5}~{\rm pc}$,
where $\epsilon_{\rm ob}$  is
the energy of the observed
photons (in ${\rm GeV}$),
 and $L_{46}$ is the
total luminosity of X--ray photons.
     At small background luminosities,
and, specifically, at small
boost factors $\delta$,
the value $r_\gamma$ may be very close to
the radius of the black hole.
       On the other hand, even if $r_\gamma > r_g$,
then near the black hole,
the X--ray radiation is expected
to be anisotropic and come from the accretion disk.
Thus, interaction of gamma--ray
photons with background radiation
may occur only at the ``walls''
of the jet, which may be
opaque to penetration of X--ray
photons (Illarionov \& Krolik 1996;
Thompson 1997).

The fastest flares reported
so far last a few hours
as in the case of Mrk 421 (Gaidos et al. 1996), down
to less than an hour in Mrk 501 (Aharonian et al. 1999).
  In PKS 1622-297 the duration
of the flare is few days,
but the rise time of the flare is only few hours
(Mattox et al. 1997).
   This indicates that the size of the emitting
region is very small (unless the Doppler boost
factor $\delta$ is not extremely large).

The Poynting flux outbursts discussed here
lead to quite small variability timescales,
from less than an hour up to possibly a few days.
The short flares are much more probable.
The frequent small--scale flares may explain the
intraday variability in different
wavebands from optics (and possibly high--frequency
radio) up to GeV and TeV radiation.

  What is
the origin of {\it the long--time flares},
which may last
a few weeks or even months?
  The long flares may be explained for example
by superposition of smaller--scale flares which may
occur frequently
(Magdziarz, Moderski \&
Madejski 1997; Chiaberge \& Ghisellini 1998).
The overlapping
of light--curves of small flares may determine the
non--thermal continuum radiation in
different wavebands from radio to gamma with  flickering
determined by the separate flares.
       The low--energy photons
have a  much longer cooling time (e.g.
Atoyan \& Aharonian 1997),
 so that the flickering
of the low--energy radiation may be smoothed significantly
(Chiaberge \& Ghisellini 1998).
         However, we still cannot exclude the possibility
 that intraday (mm) radio variability
(e.g. Wagner \& Witzel  1995) of some
 radioquasars may be also determined
by the expulsion of shocks from the region
near the black hole.

  The Poynting flux outburst
model does not exclude the
possibility of formation of longer flares
at larger distances from the black hole.
  For example, the TeV flare in Mrk 501,
may be explained
by the injection of high--energy particles with
 $\gamma_{max} \sim 10^5-10^6$ with luminosity
$L_j\sim 10^{41}~{\rm erg~ s^{-1}}$ to a region with
the size $R \sim 10^{16}~{\rm cm}$
(Maschidiadis \& Kirk 1997,
see also Sambruna et al. 1998).
  These particles may originate
in a shock wave or
a reconnection event.
   This  luminosity
 constitutes only one percent of
magnetic field annihilation
luminosity (eq. 7).

           Correlated multiwavelength
observations show
near simultaneous outbursts in
different wavebands from
 the optical
to gamma--ray (e.g., Bloom et al. 1997), which
support the Poynting flux model.
  Recent observations show clear correlation between
X-ray and TeV fluxs in TeV Blazars (e.g., Pian et al.
1998),
 which is a strong argument in favor of SSC
mechanism of TeV radiation in these
sources.
   The  TeV flares
found in our model are
determined by IC radiation, and hence cannot be
applied to these particular
sources.
  In these objects
additional acceleration of leptons
may occur in
shock waves
or by $B$ field reconnection.

\section{Conclusions}

   The main conclusions  from
the investigated models are:

1.~The rapidly rotating  inner
region of an accretion disk
threaded by magnetic field
($\sim 10^3 -10^4~\rm G$) as well
as the rapidly rotating black hole can
generate field dominated or Poynting flux
jets.

2.~Accretion disk instabilities,
specifically the ``global magnetic
instability'' of the disk,
can bring matter and magnetic flux rapidly to
the inner region of the disk and thereby
generate strong outbursts of energy to the jet.

3.~ In the region between the ``old'' and ``new'' matter
of the jet, a ``front''  forms, where particles
are efficiently accelerated owing to
shock waves and/or enhanced reconnection/annihilation
of the magnetic field.

4.~Flares in all frequency bands
from IR (or mm) to GeV (and in some cases TeV)
are expected
to appear approximately simultaneously.
  In the case where the toroidal magnetic field
reverses polarity across the front,
strong TeV flares may occur.

5.~The millimeter
radio  flux may have a short delay
(hours) compared with
other wavebands as a result of self--absorption.
 However,
in some cases (where the magnetic field
at the base of the jet
is  less than $\sim 10^3$ G)
it may appear simultaneously with other wavebands.
There is an even longer delay for lower frequency radio
emission.

6.~Frequent weak outbursts to the jet may overlap and
determine the non--thermal continuum
radiation from radio to
gamma--ray band.
They may be observed as intraday variability
in different wavebands.

7.~The strongest outbursts may appear as
short time--scale flares with duration  from less than an hour
up to a few days, depending on
 accretion processes in the disk
 and the radiation rate
of the leptons.

8.~Reconnection of magnetic field leads to
locally accelerated leptons and the possibility of
TeV flares. Small--scale reconnection events may determine
the small-scale variability at large distances
 from the black hole
with times less than $r_j(z)/c$,
where $r_j$ is the radius of the jet.

\acknowledgments

The author thanks
IAU for partial support,
 and the Local Organizing Committee
for warm hospitality.
   This work was made possible in part by
Grant No. RP1--173 of the U.S.
Civilian R\&D Foundation for the
Independent States
of the Former Soviet Union.

\medskip

\noindent{\bf Discussion}
\medskip

\noindent{\it Rita Sambruna:}
Did you apply your model to the spectral energy
distributions of blue and red Blazars?
  How do you explain
the shift in synchrotron peak frequency?
\medskip

\noindent{\it Marina Romanova:}
One possibility is that blue Blazars have
smaller viewing angles.
We plan to investigate further the
spectrum and its dependence on physical
parameters  in our future work.

\end{document}